\newcommand{\CLASSINPUTinnersidemargin}{18mm}
\newcommand{\CLASSINPUToutersidemargin}{12mm}
\newcommand{\CLASSINPUTtoptextmargin}{20mm}
\newcommand{\CLASSINPUTbottomtextmargin}{25mm}
\documentclass[10pt,conference,a4paper]{IEEEtran}

\renewcommand{\thesubsection}{\Alph{subsection}}
\renewcommand{\thesubsubsection}{\arabic{subsubsection}}

\usepackage{times}

\usepackage{graphicx}

\usepackage{amsmath}
\usepackage{times}
\usepackage{graphicx}
\usepackage{amsfonts}
\usepackage{amssymb}
\usepackage{cite}
\usepackage{subfigure}
\usepackage{url}

\DeclareGraphicsExtensions{.jpg,.eps}

\begin{document}

\title{Analysis of the electrostratic field generated by a charge distribution on a dielectric layer loading a rectangular waveguide}


\author{
\authorblockN{A. Berenguer$^{(1)}$, A. Coves$^{(1)}$, F. Mesa$^{(2)}$, E. Bronchalo$^{(1)}$, B. Gimeno$^{(3)}$ and V. Boria$^{(4)}$}
\authorblockA{aberenguer@umh.es, angela.coves@umh.es, ebronchalo@umh.es,  mesa@us.es, benito.gimeno@uv.es, vboria@dcom.upv.es}
\authorblockA{$^{(1)}$Dpto. de Ingenier\'ia de Comunicaciones. Univ. Miguel Hern\'andez de Elche. 03203, Elche (Alicante), Spain.}
\authorblockA{$^{(2)}$Dpto. de F\'isica Aplicada I. Univ. de Sevilla. 41012, Sevilla, Spain.}
\authorblockA{$^{(3)}$Dpto. de F\'isica Aplicada y Electromagnetismo-Inst. de Ciencia de Materiales. Univ. de Valencia. 46100, Valencia, Spain.}
\authorblockA{$^{(4)}$Dpto. de Comunicaciones. Univ. Polit\'ecnica de Valencia. 46022, Valencia, Spain.}
}

\maketitle

\begin{abstract}
The goal of this paper is to study the electrostatic field due to an arbitrary charge distribution on a dielectric layer in a dielectric-loaded rectangular waveguide. In order to obtain this electrostatic field, the potential due to a point charge on the dielectric layer is solved in advance. The high computational complexity of this problem requires the use of different numerical integration techniques (e.g. Filon, Gauss-Kronrod, Lobatto, ...) and interpolation methods. Using the principle of superposition, the potential due to an arbitrary charge distribution on a dielectric layer is obtained by adding the individual contribution of each point charge. Finally, a numerical differentiation of the potential is carried out to obtain the electrostatic field in the waveguide. The results of this electrostatic problem are going to be extended to model the multipactor effect, which is a problem of great interest in the space industry.
\end{abstract}

\section{Introduction}
The calculation of the electrostatic field, $E_{dc}$, in a dielectric-loaded waveguide due to an arbitrary charge distribution on the dielectric layer is a problem of great interest in the space industry, because of the lack of rigorous studies about the multipactor effect appearing in dielectric loaded waveguide-based microwave devices in satellite on-board equipment. When dealing with a partially dielectric-loaded rectangular waveguide, the electrons emitted by the dielectric surface charge the dielectric material positively, whereas the electrons absorbed by the dielectric layer charge it negatively. This charge gives rise to an electrostatic field which has to be taken into account in order to obtain an accurate trajectory of the electrons in the structure.

Lots of works have studied the electrostatic field appearing on RF dielectric windows \cite{Ang, Neuber, Kishek, Valfells, Valfells2, Yla, Anderson, Michizono}, but not so many works have studied the electrostatic field appearing during a multipactor discharge in dielectric loaded waveguides \cite{Torregrosa, Coves, Torregrosa2}.

Although the problem of obtaining the electrostatic field originated by an arbitrary electron charge distribution has been solved in many electromagnetism books \cite{Griffiths, Jackson, Bladel}, this is the first time that the problem under consideration in this work is rigorously solved, to the best of the authors' knowledge.

The paper is organized as follows: Section II describes the theory and fundamental principles underlying the problem under investigation. In Section III, the results obtained for the solution of the electrostatic problem, including three different charge distributions on the dielectric layer are shown. A few concluding remarks are made in Section IV.

\section{Theory}\label{Theory}
\begin{figure}[h!]
	\centering
		\includegraphics[width=0.45\textwidth]{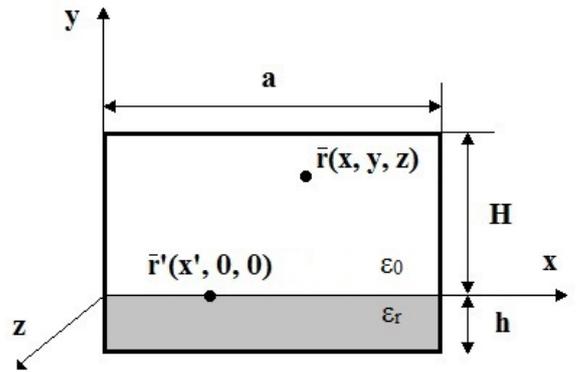}
	\caption{Geometry and dimensions of the problem under investigation.}\label{esquema}
\end{figure}

In Fig.\ref{esquema} it is shown the transverse section of the waveguide under study, consisting on a partially dielectric-loaded rectangular waveguide, whose dielectric layer has relative permittivity $\epsilon_r$ and thickness $h$.
The aim is to compute the electric field at the observation point $\vec{r} = (x,y,z)$, which is assumed to be located in the air region of a waveguide with translational symmetry along the longitudinal direction $z$, due to a point charge on the dielectric layer at $\vec{r'}=(x',0,0)$. 

In order to determine the electric field $\vec{E}(x,y,z) = -\nabla\phi(x,y,z)$, first the potential due to the point charge is calculated according to Laplace's equation for the electrostatic Green's function \cite{Collin},
\begin{equation} \label{Laplace}
	\nabla\cdot\epsilon_r(\vec{r})\nabla G(\vec{r}) = -\frac{1}{\epsilon_0}\delta(x-x')\delta(y)\delta(z)
\end{equation}
The geometric characteristics and the linear nature of the problem makes that the Dirac delta functions can be expressed as,
\begin{equation} \label{delta_x}
	\delta(x-x') = \frac{2}{a}\sum^{\infty}_{n=1}{\sin(k_{xn}x)\sin(k_{xn}x')}
\end{equation}
\begin{equation} \label{delta_z}
	\delta(z) = \frac{1}{2\pi}\int^{\infty}_{-\infty}{e^{-jk_zz}dk_z}
\end{equation}
where $k_{xn} = \frac{n\pi}{a}$. The above expressions come from the fact that the eigenfunctions of the differential operator are sinusoidal functions along $x$-axis and complex exponential functions along $z$-axis, respectively. This is equivalent to apply the discrete sine transform (DST) along $x$-axis and the integral transform along $z$-axis,
\begin{equation} \label{green_function}
	G = \frac{1}{\pi a}\int^{\infty}_{-\infty}{dk_ze^{-jk_zz}}\sum^{\infty}_{n=1}{\sin(k_{xn}x)\sin(k_{xn}x')\tilde{G}}
\end{equation}
\begin{equation} \label{green_function_tilde}
	\tilde{G} = \int^{\infty}_{-\infty}{dze^{jk_zz}}\sum^{\infty}_{n=1}{\sin(k_{xn}x)\sin(k_{xn}x')G}
\end{equation}
where $G = G(x,x',y,z)$ and $\tilde{G} = \tilde{G}(k_{xn},k_z;y)$.

According to the above considerations, Eq.(\ref{Laplace}) can be expressed as,
\begin{subequations}\label{Laplace_transformada}
\begin{equation} \label{Laplace_transformada_1}
	\left\{\frac{\partial}{\partial y} \epsilon_r(y) \frac{\partial}{\partial y} - k^2_t\right\}\tilde{G} = -\frac{\delta(y)}{\epsilon_0}
\end{equation}
\begin{equation} \label{Laplace_transformada_2}
	\tilde{G}(y=-h) = 0
\end{equation}
\begin{equation} \label{Laplace_transformada_3}
	\tilde{G}(y=H) = 0
\end{equation}
\end{subequations}
where $k^2_t = k^2_{xn}+k^2_z$. Solving Eq.(\ref{Laplace_transformada}), the following spectral Green's function, $\tilde{G}$, is obtained in the air region $y \geq 0$,
\begin{equation} \label{spectral_green_function}
	\tilde{G} = \frac{\sinh[k_t(H-y)]}{\epsilon_0 k_t[\epsilon_r \coth(k_th)+\coth(k_tH)]\sinh(k_tH)}
\end{equation}
and the Green's function, $G$, is achieved by replacing Eq.(\ref{spectral_green_function}) into Eq.(\ref{green_function}).
\begin{equation} \label{G}
	\begin{split}
	G & = \frac{2}{\epsilon_0 \pi a}\sum^{\infty}_{n=1}\sin(k_{xn}x)\sin(k_{xn}x')\\
		&\quad \times\int^{\infty}_{0}\frac{\sinh[k_t(H-y)]\cos(k_zz)}{k_t[\epsilon_r \coth(k_th)+\coth(k_tH)]\sinh(k_tH)}dk_z
	\end{split}
\end{equation}

The high computational complexity of Eq.(\ref{G}) requires the use of different numerical integration techniques (e.g., Filon, Gauss-Kronrod, Lobatto, ...). Because of the rapid oscillation of the integrand for large values of $z$	 Filon's integration method is chosen since it is desirable for integrals \cite{Hildebrand},
\begin{equation} \label{Filon}
	\int^{a}_{b}f(x)\cos(kx)dx
\end{equation}

Using superposition, the potential due to an arbitrary charge distribution on a dielectric layer is obtained by adding the individual contribution of each point charge. 
\begin{equation} \label{superposition}
	\phi (x,y,z) = \int{G(x-x',y,z)\rho(x')dx'}
\end{equation}

Finally, a numerical differentiation of the potential is carried out to obtain the electrostatic field in the waveguide.

\section{Numerical results and discussion}
This section shows the results obtained for a dielectric-loaded rectangular waveguide in Fig.\ref{esquema}. An algorithm based on the expressions given in Section \ref{Theory} has been programmed using Matlab to provide the results outlined next.

\begin{figure}[!ht]
	\centering
		\includegraphics[width=0.45\textwidth]{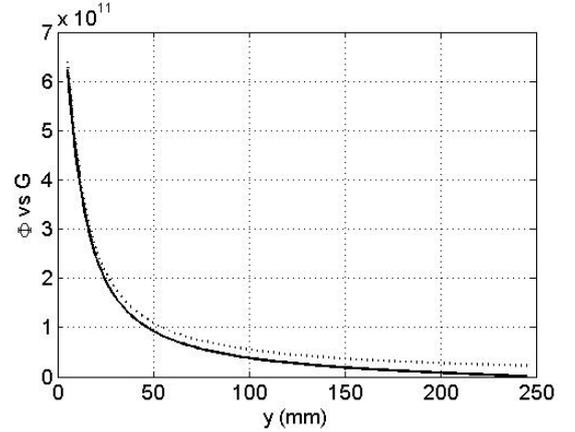}
	\caption{Comparison of the potential in the air region due to a point charge between two infinite homogeneous mediums (dashed line) vs Green's function of the problem under study (black line).}\label{comparacion_medio_inf}
\end{figure}

First, in order to validate Eq.(\ref{G}), the potential in the air region due to a point charge between two infinite homogeneous mediums ($\epsilon_{r1}=1$ and $\epsilon_{r2}$) is used as a benchmark,
\begin{equation} \label{potencial_libre}
	\phi = \frac{1}{4\pi \epsilon_0 \frac{1+\epsilon_{r2}}{2}\sqrt{(x-\frac{a}{2})^2+y^2+z^2}}
\end{equation}
The results of Eq.(\ref{G}) should approach Eq.(\ref{potencial_libre}) if the dimensions $a$, $h$ and $H$ are chosen so that the point charge and the observation point are far enough from the walls of the waveguide. In this case, the following parameters are considered: $a=600$ mm, $H=250$ mm, $h=250$ mm, $x=305$ mm, $y$ values from $5$ mm to $245$ mm with $1$ mm widthstep, $z=5$ mm, $x'=300$ mm and $\epsilon_r=2.25$. As shown in Fig.\ref{comparacion_medio_inf}, the results of Eq.(\ref{G}) and Eq.(\ref{potencial_libre}) agree as long as the observation point is far enough away from the top wall, i.e. $y=25$ mm approximately. However, beyond this $y$ value, as the observation point approaches the top wall, the approximation is no longer valid and discrepancies appear.

\begin{figure}[!ht]
	\centering
		\includegraphics[width=0.45\textwidth]{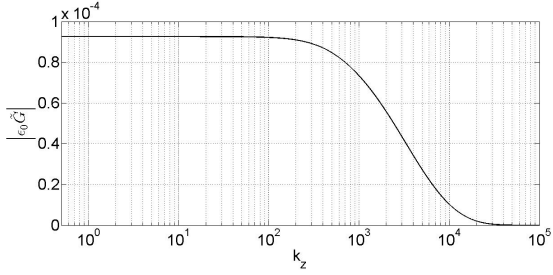}
	\caption{$\left|\epsilon_0 \tilde{G}\right|$ for the case $a=20$ mm, $H=5$ mm, $h=5$ mm, $y=0.1$ mm, $z=3$ mm, $n=1$ and $\epsilon_r=2.25$.}\label{integrando_n1}
\end{figure}

\begin{figure}[!ht]
	\centering
		\includegraphics[width=0.45\textwidth]{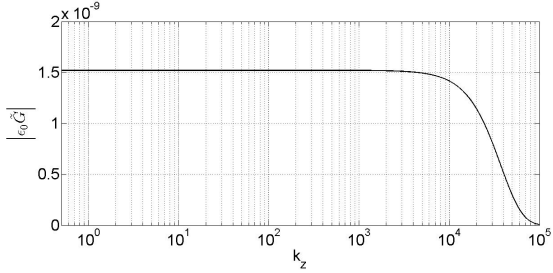}
	\caption{$\left|\epsilon_0 \tilde{G}\right|$ for the case $a=20$ mm, $H=5$ mm, $h=5$ mm, $y=0.1$ mm, $z=3$ mm, $n=500$ and $\epsilon_r=2.25$.}\label{integrando_n500}
\end{figure}

As discussed in the previous section,  the high computational complexity of this problem requires a detailed analysis of the parts forming the solution. In particular, it is useful to understand the spectral Green's function, Eq.(\ref{spectral_green_function}), with respect to the integration variable $k_z$. In this case, the following parameters are considered: $a=20$ mm, $H=5$ mm, $h=5$ mm, $z=3$ mm, $n=1$ and $\epsilon_r=2.25$. In terms of the rate of convergence, the worst scenarios are for the cases of low $y$ values and high $n$ values. In Fig.\ref{integrando_n1}, $\left|\epsilon_0 \tilde{G}\right|$ for the case of $y=0.1$ mm and $n=1$ is plotted. As it is shown, $k_z\geq2\times10^4$ has to be considered to reach convergence. $\left|\epsilon_0 \tilde{G}\right|$ for the case of $y=0.1$ mm, and $n=500$ is plotted in Fig.\ref{integrando_n500}. In this case, $k_z\geq8\times10^4$ is needed. The asymptotic behavior of the integrand, determined by the term $e^{-k_ty}$, allows us to establish a condition to stop the computation when the convergence is reached. It involves calculating the relative value of the i-th summand of the integral with respect to the accumulated value of the integral until this iteration. If this relative value is less than a particular convergence tolerance, the computation of the integral is stopped. On the other hand, regarding the convergence of the series in Eq.(\ref{G}), it depends on the product of two sinusoidal functions and no asymptotic behavior can be observed in this case. For this reason, in order to ensure that the convergence is reached, the series is decomposed into a sum of partial series of ten terms each of them. The relative value of the i-th partial series with respect to the accumulated value provides the stop condition.


\begin{figure}[!ht]
	\centering
		\includegraphics[width=0.48\textwidth]{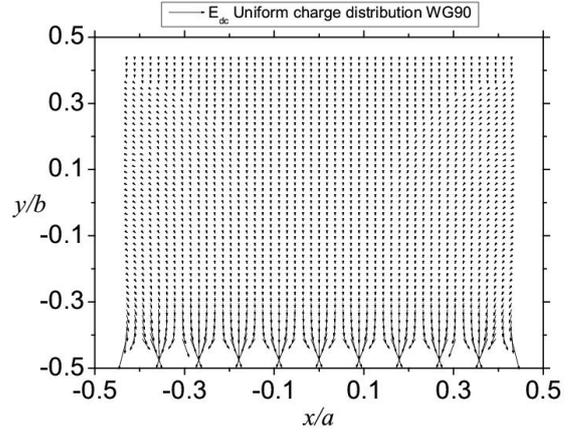}
	\caption{Electrostatic field due to an uniform charge distribution in a dielectric-loaded WG90.}\label{uniform}
\end{figure}

\begin{figure}[!ht]
	\centering
		\includegraphics[width=0.48\textwidth]{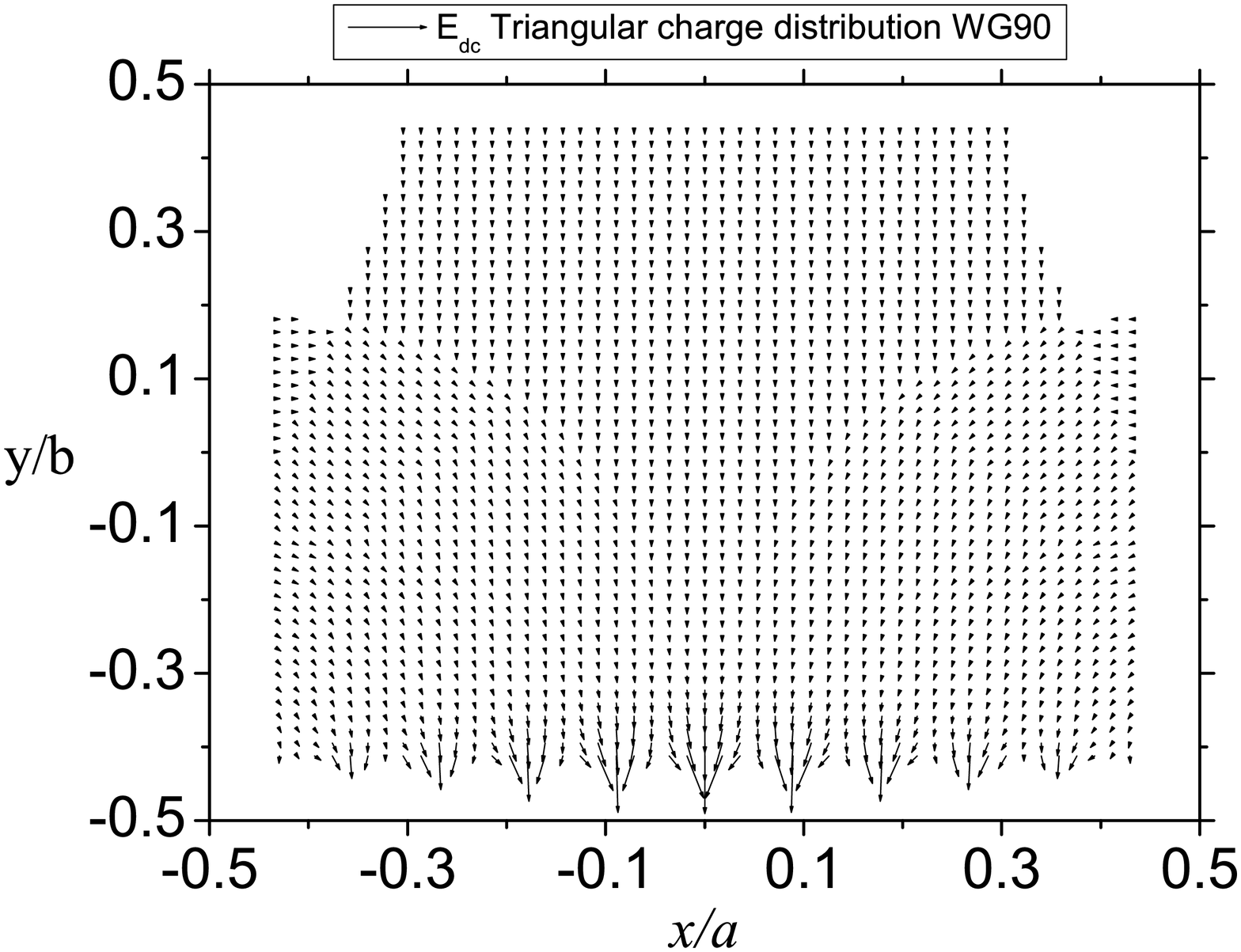}
	\caption{Electrostatic field due to a triangular charge distribution in a dielectric-loaded WG90.}\label{triangular}
\end{figure}

\begin{figure}[!ht]
	\centering
		\includegraphics[width=0.48\textwidth]{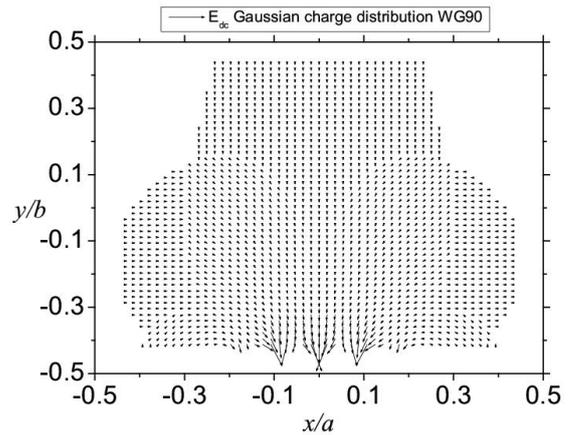}
	\caption{Electrostatic field due to a gaussian charge distribution in a dielectric-loaded WG90.}\label{gaussian}
\end{figure}

Once the solution of the electrostatic problem, Eq.(\ref{G}), has been validated and a convergence study has been carried out, the next step is to consider a problem with real dimensions. For this particular case, the parameters chosen are: $a=19.1$ mm, $H=10.135$ mm, $h=0.025$ mm and $\epsilon_r=2.1$. The origin has been located in the center of the waveguide. A total of eleven equidistant point charges has been considered in the calculation. The electrostatic field in a z-plane constant containing the point charges, $z=0$, due to an uniform charge distribution [$q_i=(1,1,1,1,1,1,1,1,1,1,1)\cdot e$], a triangular charge distribution [$q_i=(1,2,3,4,5,6,5,4,3,2,1)\cdot e$] and a gaussian charge distribution [$q_i=(0,0,0,0.2,0.8,1,0.8,0.2,0,0,0)\cdot e$] on the dielectric layer are shown in Fig.\ref{uniform}, Fig.\ref{triangular} and Fig.\ref{gaussian} respectively, where $e$ is the electron charge. As the observation points approach the dielectric layer where the point charges are located, the electrostatic field intensity becomes higher. Furthermore, as expected, a symmetrical behavior with respect to the central x-axis, $\frac{x}{a}=0$, is observed in all cases.

\section{Conclusion}
In this work, a method for calculating the electrostatic field in a dielectric-loaded waveguide due to an arbitrary charge distribution on the dielectric layer has been shown. The high computational complexity of this problem requires the use of different mathematical techniques to minimize computation time. For this purpose, it is recommendable to carry out a convergence study of the problem under study. There may be critical points, e.g. close to the walls or the dielectric layer, in the structure in which the solution does not converge properly. In these cases, it is proposed to calculate the solution at points close to them and apply extrapolation techniques.

The results of this electrostatic problem are going to be extended to model the multipactor effect, which is a problem of great interest in the space industry. 


\section*{Acknowledgment}

This work was supported by the Ministerio de Econom\'ia y Competitividad, Spanish Government, under the coordinated project TEC2013-47037-C5-4-R, TEC2013-47037-C5-1-R and TEC2013-41913-P.


\end{document}